\documentclass[onecolumn,showpacs]{revtex4}

\usepackage{graphicx}
\usepackage{dcolumn}
\usepackage{bm}

\input epsf

\begin{document}

\title{\Large A  Quintessence  Problem  in  Self-interacting
Brans-Dicke  Theory }

\author{\bf Subenoy Chakraborty}
\email{subenoyc@yahoo.co.in}
\author{\bf N. C.  Chakraborty  }
\author{\bf Ujjal Debnath}
\email{ujjaldebnath@yahoo.com}

\affiliation{Department of Mathematics, Jadavpur University,
Calcutta-32, India.}

\date{\today}

\begin{abstract}
A  quintessence  scalar  field  in  self-interacting Brans-Dicke
theory  is  shown  to give  rise to a non-decelerated  expansion
of  the  present  universe for  open, flat  and  closed models.
Along  with  providing  a non-decelerating  solution, it  can
potentially  solve  the flatness  problem  too.
\end{abstract}

\maketitle

\section{\normalsize\bf{Introduction}}
The Standard Cosmological Model (SCM) can  only  describe
decelerated  universe  models  and  so  cannot  reproduce  the
results  coming  from  the  recent  type Ia  supernovae
observations  upto  about $z \sim 1$ [1]  which  favour  an
accelerated  current  universe. But  as  the  SCM  can  give  a
satisfactory  explanation  to  other  observational  properties
of  the  present  universe  (e.g., primordial  nucleosynthesis,
extragalactic  sources  redshift, cosmic  microwave  radiation).
The  recent  extensive  search  for  a  matter  field  which can
give  rise  to  an  accelerated  expansion  for  the universe.
This  type  of  matter  field  is  called `quintessence  matter'
(or, shortly  Q-matter) This Q-matter can  behave  like  a
cosmological  constant [2] by combining positive  energy density
and  negative  pressure. So  there must  be  this Q-matter
either  neglected  or unknown responsible  for  this accelerated
universe. At  the present epoch, a  lot  of works  has  been
done  to  solve this quintessence  problem and  most  popular
candidates  for Q-matter  has  so  far  been a  scalar  field
having  a potential  which  generates  a sufficient  negative
pressure. The  quintessence  proposal faces  two  types  of
problems [3]. One  of  these  problems (referred  as  fine
tuning  problem, eliminate classically by Ratra and Peebles [2]),
is the  smallness  of the  energy  density compared  to other
typical  particle physics  scales. The other problem known  as
the  cosmic coincidence  problem  is that although the  missing
energy density  and  matter density decrease at different  rates
as the  universe  expands, it appears that the  initial condition
has  to  be  set  so precisely that the two densities become
comparable today. Quintessence  has been proposed  as that
missing  energy density  component that along  with  the matter
and baryonic density  makes the density  parameter  equal to  1.
A  special form  of quintessence  field  called  the `tracker
field' has been proposed by Ratra and Peebles to  tackle  this
problem [2] (see also ref. [4]). Since in  a variety of
inflationary  models scalar fields have been used in describing
the  transition from  the quasi-exponential expansion  of  the
early  universe to  a power  law  expansion, it  is  natural  to
try  to understand the  present acceleration  of  the  universe
which has  an exponential behaviour  too, by  constructing models
where the matter responsible  for  such  behaviour  is also
represented by  a scalar  field [5]. Inverse  power  law  is the
other potential [6]  that  has  been  studied  extensively for
quintessence models,  particularly, for  solving  the cosmic
coincidence problem. Recently, Bertolami  and  Martins [7]
obtained  an accelerated  expansion  for  the  universe in  a
modified Brans-Dicke (BD)  theory  by  introducing  a potential
which is a  function  of  BD  scalar  field  itself. Very
recently, Banerjee  et al [8,9]  also  have shown  that it  is
possible to have an  accelerated universe  with BD-theory  in
Friedmann  model  without  any matter.\\

This  paper  investigates  the  possibility  of  obtaining  a
non-decelerating  ($q \le 0$)  expansion  for  the  universe  in
BD theory  with  scalar  field  which  is  minimally  coupled to
gravity  and  serves  as  the  quintessence  matter.

\section{\normalsize\bf{Field  equations  and  solutions }}
The  Brans-Dicke  theory  is  given  by  the  action [10]

\begin{equation}
S=\int\sqrt{-g}\left[\phi
R-\frac{\omega}{\phi}\phi_{,\alpha}\phi^{,\alpha}+L_{m}\right]d^{4}x
\end{equation}

where  $\phi$  is  the  BD  scalar  field, $\omega$ is  the
dimensionless constant  BD  parameter  and  $L_{m}$  is  the
Lagrangian for all other  matter  fields. We  have  chosen  the
units   $8\pi G= c =1$.\\

The  matter  content  of  the  universe  is  composed  of
perfect  fluid  and  a  scalar  field  $\psi$ as  the
quintessence matter. We  assume that  the  universe  is
homogeneous  and  we consider  an  anisotropic  space-time  with
line-element

\begin{equation}
ds^{2}=-dt^{2}+a^{2}dx^{2}+b^{2}d\Omega_{k}^{2}
\end{equation}

where  $a, b$ are  functions  of  time  only  and

\begin{eqnarray}d\Omega_{k}^{2}= \left\{\begin{array}{lll}
dy^{2}+dz^{2}, ~~~~~~~~~~~~ \text{when} ~~~k=0 ~~~~ ( \text{Bianchi ~I ~model})\\
d\theta^{2}+sin^{2}\theta d\phi^{2}, ~~~~~ \text{when} ~~~k=+1~~
( \text{Kantowaski-Sachs~ model})\\
d\theta^{2}+sinh^{2}\theta d\phi^{2}, ~~~ \text{when} ~~~k=-1 ~~(
\text{Bianchi~ III~ model})\nonumber
\end{array}\right.
\end{eqnarray}

Here  $k$  is  the spatial  curvature  index, so  that  the
above  three  types [11] of models are Euclidean, closed  and
semi-closed  respectively.\\

Now the  BD-field  equations are

\begin{equation}
2\frac{\ddot{b}}{b}+\left(\frac{\dot{b}}{b}
\right)^{2}+\frac{k}{b^{2}}=-\frac{(p_{_{m}}+p_{_{\psi}})}{\phi}-\frac{1}{2}\omega
\left(\frac{\dot{\phi}}{\phi}\right)^{2}-2\frac{\dot{b}}{b}\frac{\dot{\phi}}{\phi}
-\frac{\ddot{\phi}}{\phi}
\end{equation}

\begin{equation}
\frac{\ddot{a}}{a}+\frac{\ddot{b}}{b}+\frac{\dot{a}}{a}\frac{\dot{b}}{b}
=-\frac{(p_{_{m}}+p_{_{\psi}})}{\phi}-\frac{1}{2}\omega
\left(\frac{\dot{\phi}}{\phi}\right)^{2}-\left(\frac{\dot{a}}{a}+\frac{\dot{b}}{b}\right)
\frac{\dot{\phi}}{\phi}-\frac{\ddot{\phi}}{\phi}
\end{equation}

\begin{equation}
\left(\frac{\dot{b}}{b}
\right)^{2}+2\frac{\dot{a}}{a}\frac{\dot{b}}{b}+\frac{k}{b^{2}}=\frac{(\rho_{_{m}}+\rho_{_{\psi}})}{\phi}-\left(\frac{\dot{a}}{a}+2\frac{\dot{b}}{b}
\right)\frac{\dot{\phi}}{\phi}+\frac{\omega}{2}\left(\frac{\dot{\phi}}{\phi}
\right)^{2}
\end{equation}

and  the  wave  equation  for  the  BD  scalar  field  $\phi$  is

\begin{equation}
\ddot{\phi}+\left(\frac{\dot{a}}{a}+2\frac{\dot{b}}{b}
\right)\dot{\phi}=\frac{1}{3+2\omega}\left[(\rho_{_{m}}-3p_{_{m}})+(\rho_{_{\psi}}-3p_{_{\psi}})
\right]
\end{equation}

$\rho_{_{m}}$  and  $p_{_{m}}$  are  the  density  and  the
pressure of normal matter, $\rho_{_{\psi}}$ and  $p_{_{\psi}}$
are  those due to the quintessence field  given  by

\begin{equation}
\rho_{_{\psi}}=\frac{1}{2}\dot{\psi}^{2}+V(\psi),~~p_{_{\psi}}=\frac{1}{2}\dot{\psi}^{2}-V(\psi)
\end{equation}

where  $V(\psi)$  is  the  relevant  potential.\\

The wave equation  for  the  quintessence  scalar  field $\psi$ is

\begin{equation}
\ddot{\psi}+\left(\frac{\dot{a}}{a}+2\frac{\dot{b}}{b}
\right)\dot{\psi}=-\frac{dV(\psi)}{d\psi}
\end{equation}

From  the  above  field  equations, we  have  the  matter
`conservation'  equation

\begin{equation}
\dot{\rho}_{m}+\left(\frac{\dot{a}}{a}+2\frac{\dot{b}}{b}
\right)(\rho_{m}+p_{m})=0
\end{equation}

Assuming  that  at  the  present  epoch, the  universe  is
filled  with  cold  matter (dust)  with  negligible  pressure,
so  using  $p_{m}=0$, the  conservation  equation (9)  gives

\begin{equation}
\rho_{m}=\frac{\rho_{1}}{ab^{2}}
\end{equation}

where  $\rho_{1}$ is  an  integration  constant.\\

Now  we  assume, the  power  law  form  of  scale  factors $a, b$
and  the  BD  scalar  field  $\phi$  are

\begin{equation}
a=a_{1}t^{\alpha},~~ b=b_{1}t^{\beta},~~ \phi=\phi_{1}t^{\delta}
\end{equation}

where  $a_{1},~ b_{1},~ \phi_{1}$ are  positive constants and
$\alpha, \beta, \delta$ are real  constants  with
$\alpha+2\beta\ge 3$ (for accelerating universe).\\

From  the  field  equations (3), (4) and (5) using (11) we have
the  expression  for $\dot{\psi}^{2}$  as

\begin{equation}
\dot{\psi}^{2}=\frac{2k\phi_{1}}{b_{1}^{2}}t^{\delta-2\beta}-\frac{\rho_{1}}{a_{1}b_{1}^{2}}
t^{-\alpha-2\beta}+\phi_{1}[2\beta^{2}-2\alpha(\alpha-1)-\omega\delta^{2}-\delta(\delta-1)
-(\alpha-2\beta)\delta]t^{\delta-2}
\end{equation}

From (6) the potential $V$ is given  by

\begin{eqnarray*}
V=\frac{k\phi_{1}}{2b_{1}^{2}}t^{\delta-2\beta}-\frac{\rho_{1}}{2a_{1}b_{1}^{2}}
t^{-\alpha-2\beta}+\frac{1}{4}\phi_{1}[(2\omega+3)(\alpha+2\beta+\delta-1)\delta+
2\beta^{2}-2\alpha(\alpha-1)
\end{eqnarray*}
\vspace{-9mm}

\begin{equation}
-\omega\delta^{2}-\delta(\delta-1)-(\alpha-2\beta)\delta]
t^{\delta-2}  \hspace{-2in}
\end{equation}

The wave equation (8) for the quintessence scalar field $\psi$
can be written in the form

\begin{equation}
-\frac{dV}{dt}=\dot{\psi}\ddot{\psi}+\left(\frac{\dot{a}}{a}+
2\frac{\dot{b}}{b}\right)\dot{\psi}^{2}
\end{equation}

After integration (14) the expression for  $V$  is

\begin{eqnarray*}
V=-\frac{k\phi_{1}}{b_{1}^{2}}\frac{(2\alpha+2\beta+\delta)}{(\delta-2\beta)}~t^{\delta-2\beta}
-\frac{\rho_{1}}{2a_{1}b_{1}^{2}}~t^{-\alpha-2\beta}+\frac{\phi_{1}(2\alpha+4\beta+\delta-2)}
{2(\delta-2)}[2\alpha(\alpha-1)
\end{eqnarray*}
\vspace{-7mm}

\begin{equation}
+\delta(\delta-1)+\omega\delta^{2}+(\alpha-2\beta)\delta-
2\beta^{2}]~t^{\delta-2}\hspace{-2in}
\end{equation}

Now, the  consistency  relations  of  the  constants  coming
from  the  two  identical  equations (13)  and (15)  for  the
potential  V  are

\begin{equation}
4\alpha+2\beta+3\delta=0
\end{equation}

\begin{eqnarray*}
(2\omega+3)\delta(\delta-2)(\alpha+2\beta+\delta-1)=(4\alpha+8\beta+3\delta-6)
[2\alpha(\alpha-1)+\delta(\delta-1)
\end{eqnarray*}
\vspace{-9mm}

\begin{equation}
+\omega\delta^{2}+(\alpha-2)\delta-2\beta^{2}] \hspace{-3in}
\end{equation}

From  these  relations  for  an  accelerating  universe
($\alpha+2\beta\ge 3$), there are  two  possibilities: one  in
which  $k = 0$ and  the second where  $k\ne 0$.\\

{\it Case I} : $ k = 0 $\\

In  this  case , consistency  conditions  are  (17)  and

\begin{equation}
\alpha+2\beta=c
\end{equation}

where  $c$  is  an  constant  ($\ge 3$).\\

For  solving  (17)  and  (18), we  may  choose  $\delta= - 2$  and
we have  two  possible  solutions:\\

(i) ~~~~ $\alpha=\beta=\frac{c}{3}$,~~~  $\delta=-2$\\

 ~~~~~~~~~~~~~~~~~~~~~~~~~~~~~ or\\

(ii) ~~~  $\alpha=4-c,~~ \beta=-2+c,~~ \delta=-2$\\

For  both  solutions  (i)  and  (ii)  the  expression  for
$\dot{\psi}^{2}$  (see  eq.(12))  becomes

\begin{equation}
\dot{\psi}^{2}=-\frac{\rho_{1}}{a_{1}b_{1}^{2}}~t^{-c}-2(2\omega+3)\phi_{1}t^{-4}
\end{equation}

This  indicates  that  $\omega<-3/2$  as  $\dot{\psi^{2}}$
cannot  be negative.\\

For  $c = 4$, one  has $2|2\omega+3|\phi_{1}\ge
\frac{\rho_{1}}{a_{1}b_{1}^{2}}$ and the deceleration parameter
$q=-\frac{1}{4}$.\\

In  this  case, equation (19) integrates  to

\begin{equation}
\psi=\pm \frac{A}{t}
\end{equation}

where
$A^{2}=-2(2\omega+3)\phi_{1}-\frac{\rho_{1}}{a_{1}b_{1}^{2}}$~~
and  the  relation  between  $V$  and  $\psi$  becomes

\begin{equation}
V=V_{1}\psi^{4}
\end{equation}

where   $V_{1}$  being  a  constant, related  to  the  constants
$a_{1}, \rho_{1}$  etc. The  model  works  for  all  time
$0<t<\infty$ with the condition  $2|2\omega+3|\phi_{1}\ge
\frac{\rho_{1}}{a_{1}b_{1}^{2}}$  is satisfied. The  value  of
$\omega$ is related to the other
constants as follows:\\

~~~~$2(2\omega+3)\phi_{1}+\frac{\rho_{1}}{a_{1}b_{1}^{2}}=
\frac{-1\pm \sqrt{1+\frac{64}{9}V_{1}}}{4V_{1}}$,~~for solution (i)\\

and\\

~~~~$2(2\omega+3)\phi_{1}=\frac{\rho_{1}}{a_{1}b_{1}^{2}}-\frac{\epsilon}{2V_{1}},
~~(\epsilon=0,1)$,~~for solution (ii)\\

For  $c \ne 4$, the  model  does  not  work  for  the  whole
range of  time $0<t<\infty$. In  this  case, the  deceleration
parameter  is  $q=\frac{3}{c}-1$.\\

If  $c>4$, then  the  rate  of  acceleration  is  faster than
$q=-\frac{1}{4}$  and  from  equation (19), $\dot{\psi}^{2}>0$
restricts the validity of the model  for  $t>t_{1}$  where\\

~~~~~~~~~~~~~~~~~~~~~~~~~~~~~~~~~~~~~~$t_{1}=\left[\frac{\rho_{1}}{2|2\omega+
3|a_{1}b_{1}^{2}\phi_{1}}\right]^{\frac{1}{c-4}}$~~~~~~~~~~~~~~~~~~~~~~~~~~~~~~~~~~~~~~~~~~($21a$)\\

Further  if  $3<c<4$, then  the  universe  will  expands with an
acceleration  but  with  a  rate  less  than  $q=-\frac{1}{4}$ and
as before from  equation (19)  for  real $\dot{\psi}$,  the  model
works upto  the time  $t_{2}$  where\\

~~~~~~~~~~~~~~~~~~~~~~~~~~~~~~~~~~~~~~$t_{2}=\left[\frac{\rho_{1}}{2|2\omega+
3|a_{1}b_{1}^{2}\phi_{1}}\right]^{\frac{1}{c-4}}$~~~~~~~~~~~~~~~~~~~~~~~~~~~~~~~~~~~~~~~~~~($21b$)\\

For  $q=-\frac{1}{4}$, the  present  age  of  the  universe  can
be calculated  from  (5)  as

$$
t_{0}=\left[2-2\omega-\frac{A^{2}}{2\phi_{1}}+\frac{V_{1}A^{4}}{\phi_{1}}\right]^{1/2}
\frac{1}{\left[(H_{b}^{2})_{0}+2(H_{a})_{0}(H_{b})_{0}\right]^{1/2}}
$$

where  $H_{a}=\frac{\dot{a}}{a}$  and  $H_{b}=\frac{\dot{b}}{b}$.\\

For  large $\omega$ limit,
$$
t_{0}\cong
\frac{\sqrt{-2\omega}}{\left[(H_{b}^{2})_{0}+2(H_{a})_{0}(H_{b})_{0}\right]^{1/2}}
$$

where $\omega$  is  obviously  a  negative  quantity.\\

{\it Case II :}  $k\ne 0$.\\

In  this  case, consistency  conditions  are  (16), (17)  and

\begin{equation}
\alpha+2\beta=3
\end{equation}

Solving  these  three  equations  we  have  the  following
solutions:\\

(i)~~~ $\alpha=1,~ \beta=1,~ \delta=-2$ ~~ for  all  values  of
$\omega$.\\

(ii)~~ $\alpha=1\mp\sqrt{\frac{2\omega+3}{2\omega-3}},~
\beta=1\pm\sqrt{\frac{2\omega+3}{2\omega-3}},~
\delta=-2\pm\sqrt{\frac{2\omega+3}{2\omega-3}} $\\

provided  for $\omega>3/2$ or $\omega\le -3/2$.\\

For  the  solution (i),  the  model  works  for  a  limited
period  of  time, $0<t<t_{1}$  where

\begin{equation}
t_{1}=2\phi_{1}\left[\frac{k}{b_{1}^{2}}-(2\omega+3)\right]\frac{a_{1}b_{1}^{2}}{\rho_{1}}
\end{equation}

For  an  open  universe  i.e., for  $k = -1, (2\omega+3)<0 $  and
$|2\omega+3|>\frac{1}{b_{1}^{2}}$.\\

For a closed  universe  i.e., for  $k = +1 ,(2\omega+3)<0 $  and
$|2\omega+3|<\frac{1}{b_{1}^{2}}$.\\

For the solution (ii) the model  works  for  the  time, where $t$
satisfies the equation

\begin{equation}
2(2\omega+3)\phi_{1}\left[\left(1\mp\sqrt{\frac{2\omega+3}{2\omega-3}}\right)^{2}-
\frac{2}{2\omega-3}\right]~t^{\pm
2\sqrt{\frac{2\omega+3}{2\omega-3}}}+\frac{\rho_{1}}{a_{1}b_{1}^{2}}~t
\le \frac{2k\phi_{1}}{b_{1}^{2}}
\end{equation}

provided  for  $\omega > 3/2$ or  $\omega\le -3/2 $.\\

\section{\normalsize\bf{Flatness  problems and its solutions }}

One important  aspect  of  this model  is  that potentially it
can solve the flatness  problem. Now we make  a  conformal
transformation [12] as

\begin{equation}
\bar{g}_{\mu\nu}=\phi~g_{\mu\nu}
\end{equation}

In  this  section, we  make the  following  transformations:

\begin{eqnarray*}
d\eta=\sqrt{\phi}~a,~~ \bar{a}=\sqrt{\phi}~a,~~
\bar{b}=\sqrt{\phi}~b,~~ \psi=\text{ln}~\phi,~~
\bar{\rho}_{_{m}}=\phi^{-2}\rho_{_{m}},~~
\bar{\rho}_{_{\psi}}=\phi^{-2}\rho_{_{\psi}},
\end{eqnarray*}
\vspace{-9mm}

\begin{equation}
\bar{p}_{_{m}}=\phi^{-2}p_{_{m}},~~\bar{p}_{_{\psi}}=\phi^{-2}p_{_{\psi}}
\hspace{-2in}
\end{equation}

So  the  field  equations  (3) - (5)  transformed  to (after some
manipulations)

\begin{equation}
\left(\frac{\bar{b}'}{\bar{b}}\right)^{2}-2\frac{\bar{a}''}{\bar{a}}-2\frac{\bar{a}'}{\bar{a}}
\frac{\bar{b}'}{\bar{b}}+\frac{k}{\bar{b}^{2}}=(\bar{p}_{_{m}}+\bar{p}_{_{\psi}})+
\frac{(3+2\omega)}{4}\left(\frac{\phi'}{\phi}\right)^{2}
\end{equation}

\begin{equation}
\left(\frac{\bar{b}'}{\bar{b}}\right)^{2}+2\frac{\bar{a}'}{\bar{a}}\frac{\bar{b}'}{\bar{b}}+
\frac{k}{\bar{b}^{^{2}}}=(\bar{\rho}_{_{m}}+\bar{\rho}_{_{\psi}})+
\frac{(3+2\omega)}{4}\left(\frac{\phi'}{\phi}\right)^{2}
\end{equation}

and

\begin{equation}
\frac{\bar{b}''}{\bar{b}}-\frac{\bar{a}''}{\bar{a}}+\frac{k}{\bar{b}^{2}}=
\frac{\bar{a}'}{\bar{a}}\frac{\bar{b}'}{\bar{b}}-\left(\frac{\bar{b}'}{\bar{b}}\right)^{2}
\end{equation}

where~~  $' \equiv \frac{d}{d\eta}$.\\

The BD scalar field in new version $\bar{\rho}_{_{\phi}}$ is given
by

\begin{equation}
\bar{\rho}_{_{\phi}}=\frac{(3+2\omega)}{4}\left(\frac{\phi'}{\phi}\right)^{2}=\bar{p}_{_{\phi}}
\end{equation}

We define the dimensionless density parameter $\bar{\Omega}$ as

\begin{equation}
\bar{\Omega}=\bar{\Omega}_{_{m}}+\bar{\Omega}_{_{\phi}}+\bar{\Omega}_{_{\psi}}=
\frac{\bar{\rho}}{3\bar{H}^{2}}
\end{equation}

where
$\bar{\rho}=\bar{\rho}_{_{m}}+\bar{\rho}_{_{\phi}}+\bar{\rho}_{_{\psi}}$
is the total
density and $\bar{\Omega}_{i}$ are defined accordingly.\\

Using (27)-(30), and combining the energy densities, we have the
equation for the conservation for the total energy,

\begin{equation}
\bar{\rho}'+3\bar{H}(\bar{\rho}+\bar{p})=0
\end{equation}

Here
$\bar{H}=\frac{1}{3}\left(\frac{\bar{a}'}{\bar{a}}+2\frac{\bar{b}'}{\bar{b}}\right)$
is  the  Hubble  parameter  in  the Einstein frame and  $\gamma$
is the  net  barotropic  index  defined  as
\begin{equation}
\gamma~\bar{\Omega}=\gamma_{_{m}}~\bar{\Omega}_{_{m}}+\gamma_{_{\phi}}~\bar{\Omega}_{_{\phi}}
+\gamma_{_{\psi}}~\bar{\Omega}_{_{\psi}}
\end{equation}

From  equations (28)  and  (32), we  have  the  evolution
equation  for  the  density  parameter  as

\begin{equation}
\bar{\Omega}'=\bar{\Omega}(\bar{\Omega}-1)[\gamma\bar{H}_{_{a}}+2(\gamma-1)\bar{H}_{_{b}}]
\end{equation}

where ~~ $\bar{H}_{_{a}}=\frac{\bar{a}'}{\bar{a}}$    and
$\bar{H}_{_{b}}=\frac{\bar{b}'}{\bar{b}}$ .\\

The individual $\gamma_{i}$~'s are defined by the relation
$p_{i}=(\gamma_{i}-1)\rho_{i}$. So the ratios
$\frac{p_{i}}{\rho_{i}}$ remain same in both frames. For our
choices of matter, $\bar{p}_{_{m}}$ and
$\bar{p}_{_{\psi}}=\bar{\rho}_{_{\psi}}$, so we have
$\gamma_{_{m}}=1$ and $\gamma_{_{\phi}}=2$. The other index
$\gamma_{_{\psi}}$ is related by the equation

\begin{equation}
\gamma_{_{\psi}}=\frac{p_{_{\psi}}+\rho_{_{\psi}}}{\rho_{_{\psi}}}=\frac{\dot{\psi}^{2}}
{\frac{1}{2}\dot{\psi}^{2}+V}
\end{equation}

It has been shown that $\gamma_{_{\psi}}$ is varies with time.
The equation (34) indicates that $\bar{\Omega}=1$ is a possible
solution and this solution determines that
$\left(\frac{\partial\dot{\bar{\Omega}}}{\partial\bar{\Omega}}\right)_{\bar{H}}<0$.
This solution is stable for expanding universe $(\bar{H}>0)$ if
$\gamma<\frac{2}{3}$ with the relevant condition

\begin{equation}
\bar{\Omega}_{_{m}}+4\bar{\Omega}_{_{\phi}}<(2-3\gamma_{_{\psi}})\bar{\Omega}_{_{\psi}}
\end{equation}

From  the  field equation  (28), the  curvature  parameter
$\bar{\Omega}_{_{k}}=-k /~\bar{b}^{2}$ vanishes  for  the
solution $\bar{\Omega}=1$. So for BD-scalar field it  is possible
to have a stable  solution corresponding  to $\bar{\Omega}=1$ and
hence  the flatness problem  can  be solved.

\section{\normalsize\bf{Concluding  remarks }}
In  this  work, we  have  investigated  the  nature  of  the
potential  relevant  to  the  power  law expansion  of  the
universe  in  a  self-interacting  Brans-Dicke (BD) cosmology
with  a  perfect  fluid  distribution  for  anisotropic
cosmological  models. We  have  considered  a non-gravitational
quintessence  scalar  field  $\psi$  with  a potential $V =
V(\psi)$. This scalar  field  in  BD-theory  is shown  to give
rise  to an accelerated  expansion  for  the present  dust
universe (where we have taken $p_{m}=0$) It  is  to  be noted
that at early stages of  the  evolution  of  the universe $p_{m}$
is non-zero. But if  we take  barotropic  equation of  state
$p_{m}=(\gamma-1)\rho_{m}$, then equation (10) is modified  to
$\rho_{m}=\frac{\rho_{1}}{(ab^{2})^{\gamma}}$.\\

We  have  presented  a  class  of  solutions  describing
non-decelerating  universe  for  both  flat ($k=0$) and  curved
space-time ($k\ne 0$). For  $k=0$, there  are  two  possible
solutions for  different choice  of  the  parameters. In  both
the  solutions, the  parameter $\omega$ must  be  negative (in
fact $2\omega+3$ is  negative) to  make  the  quintessence  field
real. The validity (on  the  time  scale) of  the  solutions
depends on the  parameter  $c$  (defined  in  eq.(18)). For
$c=4$, the model  works  for  all  time $0<t<\infty$, while  for
$c>4$, the model works  for ($t_{1}, \infty$) where $t_{1}$ is
given by ($21a$). From this model  we  cannot  predict  the
geometry  of the universe before $t_{1}$. Similarly  for  $3<c<4$
the  model  is valid for  $t<t_{2}$  i.e., from  the  begining  to
the time $t_{2}$. Here also we cannot  predict  the  nature  of
space-time of  the universe after $t=t_{2}$.\\

In  curved  space-time ($k\ne 0$)  there  are  also  two  possible
solutions  for  different  choice  of  the  parameters.  The
time  interval  over  which  the  solutions  are  valid  are
given  in  equations (23)  and  (24). For  close  model ($k>0$)
the  coupling  parameter  $\omega$  is  restricted  by  the
inequality $0<2\omega+3<1/b_{1}^{2}$ while   for   open  model
$\omega$ satisfies $-1/b_{1}^{2}<2\omega+3<0$. From   the   above
solutions, we note that the coupling parameter  $\omega$  may  be
positive  or negative (with some restrictions). Hence  for  all
solutions the choice of $\omega$ is not in  agreement  with  the
observations.\\

Finally, for  non-decelerating  solution  it  can  also
potentially  solve  the  flatness  problem (without  any
restriction  on  the  parameters)  and  it  has  been  shown that
$\bar{\Omega}=1$ could  be  a  stable  solution  in  this  model.\\

{\bf Acknowledgement:}\\

The  authors  are  thankful  to  the  Relativity  and  Cosmology
Research  Center, Department  of  Physics, Jadavpur  University
for  helpful  discussion. One  of  the authors (U.D)  is
thankful  to  CSIR (Govt. of India)  for  awarding  a  Junior
Research Fellowship.\\

{\bf References:}\\
\\
$[1]$  Perlmutter  S  et al {\it Astrophys.~J.} {\bf 517} 565
(1999); Riess  A  G  et al, {\it Astron.~J.} {\bf 116} 1009
(1998); Garnavich P. M. et al, {\it   Astrophys.~J.} {\bf 509} 74
(1998); G. Efstathiou et al,  {\it astro-ph}/9812226.\\
$[2]$  B. Ratra  and  P. J. E. Peebles, {\it Phys. Rev. D} {\bf
37} 3406 (1988); See also R. R. Caldwell, R. Dave and  P. J.
Steinhardt, {\it Phys.
Rev. Lett.} {\bf 80} 1582 (1998) [{\it astro-ph}/9708069].\\
$[3]$  P. J. Steinhardt, L. Wang  and  I. Zlatev, {\it Phys. Rev. Lett.}
 {\bf 59} 123504 (1999).\\
$[4]$  I. Zlatev, L. Wang and P. J. Steinhardt, {\it Phys. Rev. Lett.} {\bf 82} 896 (1999).\\
$[5]$  A. A. Strarobinsky, {\it JEPT  Lett.} {\bf 68} 757 (1998);
T. D. Saini, S. Roychaudhury,  V. Sahni, A. A. Strarobinsky, {\it
Phys. Rev. Lett.} {\bf 85} 1162 (2000).\\
$[6]$  P. J. E. Peebles and B. Ratra, {\it Astrophys. J. Lett.}
{\bf 325} L17 (1998);  P. G. Ferreira  and  M. Joyce, {\it Phys.
Rev. Lett.} {\bf 79} 4740 (1987); E. J. Copeland, A. R. Liddle
and  D. Wands, {\it Phys. Rev. D} {\bf 57}  4686 (1988).\\
$[7]$  Bertolami O and  Martins  P  J {\it Phys. Rev. D} {\bf 61}
064007 (2000); Chimento  L  P, Jakubi A  S  and  Pavo'n  D {\it
Phys. Rev. D} {\bf 62}  063508 (2000).\\
$[8]$  Banerjee  N  and Pavo'n D  {\it Phys. Rev. D} {\bf 63}
043504  (2001).\\
$[9]$  Banerjee  N  and Pavo'n D  {\it Class. Quantum Grav.} {\bf
18} 593 (2001).\\
$[10]$  C. Brans and  R. H. Dicke, {\it Phys. Rev.} {\bf 124} 925
(1961).\\
$[11]$  Thorne K S {\it Astrophys.~J.} {\bf 148} 51 (1967).\\
$[12]$  Faraoni V, Gunzig  E and Nardone P  {\it Fundam. Cosm.
Phys.} {\bf 20} 121 (1999).\\

\end{document}